\documentclass[apj]{emulateapj}  

\usepackage{graphicx}
\usepackage{natbib}

\newcommand{\beq}{\begin{equation}}
\newcommand{\eeq}{\end{equation}}

\newcommand{\etal}  {{\it et al.}}

\newcommand{\Lower}[1]{\smash{\lower 1.5ex \hbox{#1}}}
\def\etal{{\it et al.\thinspace}}

\usepackage{morefloats}  

\slugcomment{Submitted to the Astrophysical Journal  5 May 2016} 

\shorttitle{Kepler Exoplanets}
\shortauthors{Traub}

\begin{document}


\title{Kepler Exoplanets: A New Method of Population Analysis}


\author{Wesley A. Traub }
\affil{Jet Propulsion Laboratory, California Institute of Technology, Pasadena, CA 91109}
\email{wtraub@jpl.nasa.gov}



\begin{abstract}
This paper introduces a new method of inferring the intrinsic exoplanet population from Kepler data, based on the assumption that the frequency of exoplanets can be represented by a smooth function of planet radius and period.  The method is applied to the two most recent data releases from the Kepler project, q1-16 and q1-17, over the range of periods 0.5 to 512 days, and radii 0.5 to 16 Earth radii.  Both of these releases have known biases, with the first believed to contain excess false positives, and the second excess false negatives, so any analysis of them should be viewed with caution.  We apply the new method of population estimation to these releases, treating them like practice data sets.  With this method, we tentatively find that the average number of planets per star would be about $5.7\pm0.8$ for F stars, $5.0\pm0.2$ for G stars, $4.0\pm0.3$ for K stars, and $6.5\pm1.7$ for M stars, indicating a decreasing trend with FGK spectral type, but an upward jump for M stars.  A second conclusion is that the number of planets per G star, per natural log unit of period (days) and radii (Earths) at the period and radius of the Earth around the Sun, is about $\Gamma_\oplus(G) = 1.1\pm 0.1$.  A related parameter, $\eta_{\oplus}$, which in addition depends on the range of period and radius considered, is found to be $\eta_{\oplus}(G) \simeq 1.0 \pm 0.1 $.  More definitive conclusions, and validation of these preliminary values, await the final release of Kepler's transiting exoplanet list.
\end{abstract}



\section{Introduction} \label{intro}

Knowledge of the intrinsic population of exoplanets serves many purposes: to predict the expected yields of future exoplanet missions; to compare with existing catalogs of exoplanets obtained by other techniques including radial velocity, gravitational microlensing, and direct imaging; and to serve as a boundary condition for theories of planet formation and evolution.

The Kepler mission's transiting exoplanet data provide a window into the intrinsic population of exoplanets around Kepler target stars.  In practice, this window is not as clear as we would wish, owing to uncertainty in extracting the signature of transiting exoplanets from a noisy data stream.

There are several recent review papers regarding the Kepler mission and the population of exoplanets that can be inferred from the Kepler data, e.g.,
\citet{Batalha2014};
\citet{Borucki2016}; and
\citet{Winn2015}.
Each of these contains relevant background information used in the present paper.

There are many published analyses and related studies of the Kepler data for the purpose of estimating the underlying number of planets per star in the target star population, e.g., (by year and author)
\citet{Dressing2015};    
\citet{Mulders2015};    
\citet{Mullally2015};    
\cite{Rogers2015};
\citet{Rowe2015};       
\citet{Silburt2015};    
\citet{Farr2014};        
\cite{Foreman2014};       
\cite{Kane2014};          
\cite{Beauge2013};       
\citet{Berta2013};      
\citet{Christiansen2015};    
\citet{Dressing2013};      
\citet{Dong2013};           
\citet{Fressin2013};         
\citet{Morton2014};           
\citet{Petigura2013a};       
\citet{Petigura2013b};        
\citet{Swift2013};           
\citet{Gaidos2012};           
\citet{Howard2012};          
\citet{Mann2012};            
\citet{Traub2012};            
\citet{Tremaine2012};          
\citet{Catanzarite2011};       
\citet{Youdin2011};           
and
\citet{Howard2010};          

Detection efficiency can be estimated by three different methods, as pointed out by \citet{Foreman2014}:
(1) assuming that the catalog is complete (e.g., \citet{Catanzarite2011}; \citet{Traub2012}; \citet{Tremaine2012});
(2) assuming an analytic form for the detection efficiency as a function of signal-to-noise (e.g., \citet{Youdin2011}; \citet{Howard2012}; \citet{Dressing2013}; \citet{Dong2013}; \citet{Fressin2013}; \citet{Morton2014}); and
(3) determining the detection efficiency by injecting synthetic signals into the raw data and testing recovery (e.g., \citet{Christiansen2015}; \citet{Petigura2013a}, \citet{Petigura2013b}).  We use  method (2) in this paper.

Occurrence rates can be estimated by these methods (also from \citet{Foreman2014}):
(A) the \emph{inverse-detection efficiency} method, which has been used by many authors; and
(B) the \emph{likelihood function} method, which is used in the Foreman-Mackey paper.
We note that there is a third method of inferring the population:
(C) the \emph{forward modeling} method, which starts with a parameterized population model plus the detection efficiency, and fits this to the observed distribution of planets.  We use method (C) in this paper.

The surprising result of all the thought that has gone into estimating occurrence rates is that there is very little agreement regarding the results.  As a key example, we note that \citet{Foreman2014} derive a value for the occurrence rate of planets around Sun-like stars, per natural logarithmic unit of period (days) as well as planet radius (Earth radii), $\Gamma_\oplus$, in the neighborhood of the Earth's period and radius (defined in Sec.~\ref{gamma-earth}), that is a factor of 6 times \textit{smaller} than that of \citet{Petigura2013b}, despite using the same catalog of planets and the same completeness function.  On the other hand, the present paper finds a value for this quantity that is about 9 times \textit{larger} than Petigura's (Sec.~\ref{gamma-earth}).

Neither \citet{Foreman2014} nor the present author have an explanation for this apparent disagreement in results.  One can look to the differences in data bases and differences in methods, but none of these provide an obvious answer.  Therefore the present paper should be seen as a presentation of a method of analysis, but not necessarily of specific numerical results.  We give sufficient detail in the model that it can be reproduced elsewhere with a minimal investment of time.  Furthermore the data bases in the present paper have known flaws, which provides a further reason for the numerical results to be given less weight than the method of analysis.  It is expected that the final results of the Kepler mission for numbers of transiting planets will become available soon, so the present work is directed toward analyzing that final data set.

In this paper we outline the method of analysis in Sec.~\ref{method},  set up the databases of Kepler's target stars and observed planets in Sec.~\ref{databases}, and describe the contents of these databases in Sec.~\ref{obs}.  The instrument model is described in Sec.~\ref{inst}.  We motivate the use of power laws and show the result for a one-segment power law in Sec.~\ref{prep}.  We treat the case of a two-segment power law in Sec.~\ref{broken}, and show that it can account for most of the major features of the observed data sets, with some notable exceptions.
We present estimated occurrence rates as a function of spectral type (Sec.~\ref{occ-spty}), occurrence parameters (Sec.~\ref{occ-param}), and Earth-like properties (Sec.~\ref{occ-eta123}). The paper concludes with a discussion (Sec.~\ref{summary}) regarding the application of the type of model proposed here to the expected final exoplanet data release of the Kepler project.  Appendix A discusses the influence of limb-darkening on the transit signal.  Appendix B provides an explicit normalization for Eqn.~\ref{eqn-h}.

\section{Method of analysis} \label{method}

We introduce and explore a new method of inferring the intrinsic population of exoplanets using a representation in terms of planet radius, period, and host star spectral type.  The representation explored is a simple power law in radius and period, which, as we will show, is suggested by the data itself; other functional forms may be found to work better, but for this initial inquiry a power law appears to be adequate.  The method of inference uses a direct fit of a functional form to the full range of observed data, which avoids potential uncertainties in alternative methods that focus on smaller segments of the observed data.  The method is outlined as follows.

Along with many other authors, we assume that the average number of planets per star $N(p,r)$ as a function of planet period $p$ (days) and radius $r$ (Earth radii) can be represented by an equation of the form
\begin{equation}
  \frac{d^2N(p,r)}{d\ln(p) \cdot d\ln(r)} =   \overline{N} \cdot h(p,r).
  \label{eqn-d2n}
\end{equation}
Here $\overline{N}$ is a scaling factor and $h(p,r)$ is a shape factor normalized to unity over the range of period $(p_{min}, p_{max})$ and radius $(r_{min}, r_{max})$ under consideration and for which observed data is available, so that we require
\begin{equation}
  \int_{p_{min}}^{p_{max}} \int_{r_{min}}^{r_{max}} h(p,r) \cdot d\ln(p) \cdot d\ln(r) = 1.
  \label{eqn-h}
\end{equation}
The shape factor $h(p, r)$ is also understood to be a function of host star spectral type, although for simplicity this is not usually called out.  The scaling factor $\overline{N}$ is thus seen to be the average number of planets per star within the (min., max.) ranges of $p$ and $r$.

The variables $p$ and $r$ are continuous, but for the sake of working with observed data it is desirable to group data into discrete intervals, ``bins", with central or average values given by $p_i$ and $r_j$, where $i$ and $j$ are bin index numbers, as detailed below.

Let $N_{obs}(p_i,r_j)$ be the observed number of planets in the $i$-th period bin centered at $p_i$(days), and the $j$-th radius bin centered at $r_j$(Earth radii).  We use a range for $i$ and $j$ of 1 to $i_{max} = j_{max} = 20$.   Note that $N_{obs}$ is implicitly a function of host star spectral type (SpTy).  Because the distribution of observed planets is locally more uniform in $[ ln(p),ln(r) ]$ space than in $(p,r)$ space, the bins are chosen to be of constant size in logarithmic coordinates.

The present method of analysis has two phases. In the first phase we assume that each target star $s_k$ has exactly one planet around it, and we simulate the expected number $N_1(p_i, r_j)$ of Kepler-detected planets in each $(p_i, r_j)$ bin.  We also assume that this planet is equally likely to be in any of the $i_{max} \cdot j_{max} = 400$ available bins, so the average number of planets per star is $n_0 \equiv 1/400$ in each bin.  Thus $h(p_i, r_j) = h_1 = const.$, subject to normalization (Eqn.~\ref{eqn-h}).  Explicitly we have
\begin{equation}
  N_1(p_i, r_j) =  \sum_k  n_0 \cdot P_{det}(p_i, r_j, s_k)
  \label{eqn-n1}
\end{equation}
where $P_{det}(p_i, r_j, s_k)$ is the probability of detecting a planet with period $p_i$ and radius $r_j$ around star $s_k$, using Kepler, and the sum is over all target stars of the desired spectral type.  See Sec.~\ref{inst} for details of calculating $P_{det}$, but in short we note here that $P_{det}$ is non-zero only for those planets that transit their star and produce a strong enough transit signal as well as a sufficient number of transits to give a mission-length signal-to-noise ratio (SNR) that yields a finite probability of detection.  Note that for a given instrument configuration and set of target stars, $N_1$ is a fixed array.

In the second phase we compare the observed number of planets per bin $N_{obs}(p_i, r_j)$ to a simulated number $N_{sim}(p_i, r_j)$, where in the simulation we iteratively adjust the overall value of $\overline{N}$ and the relative shape of $h(p_i, r_j)$ across all bins, again subject to normalization, until the observed and simulated number of planets per bin are well-matched in a least-squares sense, over the full range of period and radius, i.e., over all 400 bins.  Explicitly, $N_{sim}$ is given by
\begin{equation}
  N_{sim}(p_i, r_j) =  \overline{N} \cdot N_1(p_i,  r_j) \cdot h(p_i, r_j)
  \label{eqn-nsim}
\end{equation}
and the $h(p_i,r_j)$ term is to be adjusted by varying the internal parameters of $h$ (see Sec.~\ref{prep} and \ref{broken}).  Note that $N_1$ here is the same as above (Eqn.~\ref{eqn-n1}), so we use the ``one-planet-per-star" concept twice in this method of analysis.

We define the chi-square estimate of distance between the observed and simulated sets of values as
\begin{equation}
  \chi^2(\overrightarrow{v}) = \sum_i \sum_j ( N_{obs}(p_i, r_j) - N_{sim}(p_i, r_j, \overrightarrow{v}) )^2
  \label{eqn-chisq1}
\end{equation}
where the sum is over each bin of $p_i$ and $r_j$, and $\overrightarrow{v}$ is the vector of parameters $\overline{N}$ plus those that govern the shape factor $h(p,r)$.

We carry out the least-squares fit
\begin{equation}
  \partial \chi^2 / \partial \overrightarrow{(v)} = 0
  \label{eqn-chisq2}
\end{equation}
to estimate the shape of $h(p,r)$.  Then with $h$ known we simulate the number of planets per star in any range as
\begin{equation}
  N_{range} = \overline{N} \cdot    \int \int h(p,r) \cdot d\ln(p) \cdot d\ln(r),
  \label{eqn-n}
\end{equation}
where the integrals are taken over the desired ranges of $p$ and $r$, which can be smaller or greater than the (min., max.) input data ranges of $p$ and $r$.

The advantage of this approach is that by fitting a smooth function to the numbers of observed planets (from 0 up to about 50) in each bin, we are using an intrinsically well-controlled process.  At no point do we divide the observed number of planets by a small number to invert the observations to give the parent distribution, which can be unstable owing to the smallness of the divisor in places where the instrumental efficiency is very small (long periods and small planets), and owing as well to the small number of observed planets per bin (often 0 or 1 here).

\begin{deluxetable}{ccccccc}  
\tabletypesize{\small}
\tablecaption{Star - Planet Data Ranges   \label{table-star-planet}}
\tablewidth{0pt}
\tablehead{
   SpTy & $T_{min}$ & $T_{max}$  & $N_{star}$ & $N_{planet}$ & $N_{star}$ & $N_{planet}$ \\
        & (K)       &    (K)     &   q1-16   &  q1-16      &    q1-17  &  q1-17     }
\startdata
   FGK  &  3900   & 7300  & 155,693   &  4161 & 156,116  & 3446  \\
    F   &  6000   & 7300  &  62,261   &  1463 & 62,333   & 1143  \\
    G   &  5300   & 6000  &  64,517   &  1825 & 64,758   & 1539  \\
    K   &  3900   & 5300  &  28,915   &  873  & 29,025   &  764  \\
    M   &  2400   & 3900  &   3,683   &  167  &  3,674   &  163  \\
  all   &         &       & 194,873   &  5622 & 195,359  & 4235
\enddata 
\end{deluxetable}  

\section{Star and Planet Databases} \label{databases}

We start with the two most recent catalogs of stars and exoplanets from the Kepler mission.  We caution that each of the exoplanet catalogs in this paper was generated using different pipelines and other criteria which are known to have generic flaws, so they should be considered as practice data sets.  Thus any derived science parameters should be viewed with this caution in mind.

The exoplanet catalogs are downloaded from exoplanetarchive/ipac/caltech.edu by selecting ``data" then ``KOI (all lists)" then ``Q1-Q16 Done" which we label as ``q1-16", and ``Q1-Q17 DR24 Done" which we label as ``q1-17".  Note that q1-17 is called data release (DR) 24, to distinguish it from the expected final analysis of all 17 quarters which is expected to be labeled as DR 25.

For exoplanets in q1-16 we remove false positives and entries with blank planet periods and radii.
The remaining list contains 997 confirmed planets, 1052 candidates, and 3573 ``not dispositioned", for a total of 5622 entries.

For q1-17 we remove false positives, and entries with blank planet periods and radii, blank J magnitudes, and blank star radius errors.  The remaining list contains 974 confirmed and 3261 candidates, for a total of 4235 entries.

The target star catalogs are downloaded from the same site, by going to ``data" then ``Kepler stellar".  The 194,873 entries labeled ``q1-q16stellar" are saved for use with the q1-16 exoplanet data.  The 195,359 entries labeled ``q1-q17-dr24-stellar" are saved for use with the q1-17 exoplanet data.

The total numbers of exoplanets and stars in the data sets are listed in Table~\ref{table-star-planet} by spectral type of target star.  The effective temperature range of each spectral type is adapted from \citet{Pecaut2013}.  The row labeled ``all" indicates the total number of stars or planets in each database, which includes targets with effective temperatures outside the range of 2400 to 7300 K, as well as targets with $log(g)$ values outside the main-sequence range.

\begin{figure}[htbp]
\centering
\includegraphics[width=3.4in]{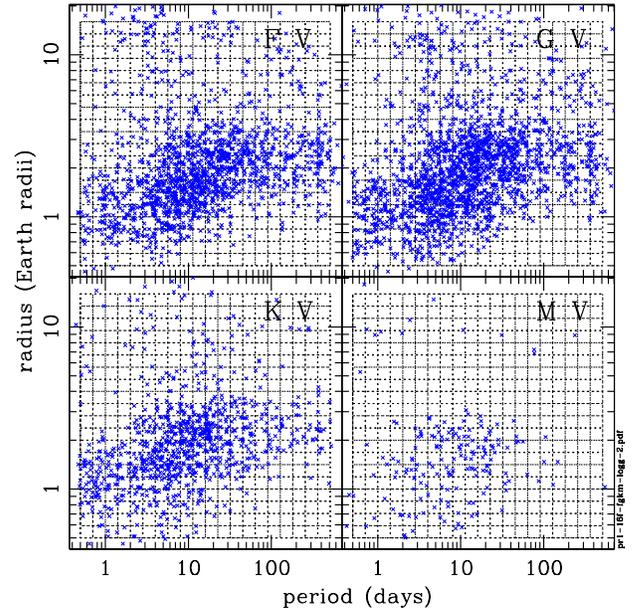}
\caption{ The period (days) and radius (Earth radii) are shown for each of the confirmed and candidate exoplanets in the q1-16 Kepler database in this paper. The plot for q1-17 (not shown) is similar but with fewer planets in the period ranges 0.5 to 2.0 days, as well as 100 to 500 days.   In this paper, the $20\times20$ grid of 400 bins is the region of $(p,r)$ space over which the observed number of planets are compared with the simulated numbers.}
 \label{fig-input-planets}
\end{figure}

\section{The Observed Data}\label{obs}

The q1-16 transit search used stellar targets \citep{Huber2014a}, a planet search algorithm \citep{Tenenbaum2014}, SOC pipeline version 9.1 \citep{Huber2014a}, and an estimated efficiency of detection by the pipeline \citep{Christiansen2015}.  The q1-17 transit search used stellar targets \citep{Huber2014b}, a planet search algorithm (\citet{Seader2015} and \citet{Coughlin2015}), SOC pipeline version 9.2 \citep{Huber2014b}, and a detection efficiency estimate \citep{Christiansen2016}. The q1-16 planets are shown in Fig.~\ref{fig-input-planets}.

The stellar and planet data in both searches are broadly similar, with an important exception noted at the end of this section.  The data available from both searches, and used in the present paper, are described in this section.

The q1-16 stellar data are discussed in \citet{Huber2014a}, who quote a typical uncertainty of $\sim40\%$ for radius and $\sim20\%$ for mass, for stars with photometric constraints, and $5-15\%$ in radius and $\sim10\%$ in mass for stars based on spectroscopy and/or asteroseismology. For q1-17 typical uncertainties in stellar mass and radius are shown in Fig.~2 in \citet{Huber2014b}.   We use the effective temperature to assign a star to a spectral type, and its surface gravity to select main-sequence dwarf-type stars, in the $log(g)$ range $4.0-5.41$ (see Fig.~1 in \citet{Huber2014b}).  Mass is used to convert planet period to semi-major axis, and along with star radius is used to calculate transit probability.

In the stellar databases ``dataspan" is the time elapsed between the first and last cadences containing valid data, with range about 21 to 1426 days, with a median of 1426 days;  a star's dataspan value is used to calculate whether the required minimum number of 3 transits could have been observed.

Likewise ``dutycycle" is the fraction of data cadences within the span of observations that contain valid data and contribute toward detection of transit signals; the range of dutycycle values is from about 0.09 to 0.99, with a median of about 0.88.  Dutycycle is not used in this paper but could be used in a more refined estimate of the likelihood of detection of long-period planets.

The multiple-event statistic (MES) threshold is given for each target star for transit durations of 1.5 to 15.0 hours, in steps of 0.5 to 2.5 hours.  A value of 7.1 indicates that the transit planet search (TPS) module reached the nominal search threshold significance, whereas a value above 7.1 indicates that TPS ended the search prematurely at the higher specified significance threshold. The MES is essentially the same as the mission signal to noise ratio (SNR) for detection of a transiting planet, and the terms will be used interchangeably in this paper.   The tabulated MES threshold is used here, interpolated to the model transit time, to decide whether a model planet transiting a target star would result in a valid detection.

The combined differential photometric precision (CDPP) is an empirical estimate of the noise in the relative flux time series observations.  TPS estimates a non-stationary time series of CDPP, which sets the significance level of transit signals detected in the flux time series.  The CDPP is used here to calculate the expected MES of a model transit.

For a given expected MES of a model transit, the expected detection efficiency (probability of detection) has been estimated by \citet{Christiansen2015} for the q1-16 pipeline.  The efficiency was estimated by injecting simulated planet signals into the pixel data of about 10,000 target stars spanning one year of observations, and processing the time series through the pipeline.  They find that the resulting sensitivity curve (fraction of transits detected) is near zero for MES values less than about 5, and near unity for MES values greater than about 15, with a half-point at about 8.5, well above the detection threshold (7.1) in the pipeline.  The resulting empirical curve of the fraction detected can be well characterized by a $\Gamma$ cumulative distribution function, and they give the parameters for this function for the q1-16 pipeline, including the MES offset value needed for the class of FGK stars.  The gamma function parameters for q1-17 are given by \citet{Christiansen2016}, and listed in Table~\ref{table-fdet}.  It is found that the pipeline produces a different set of parameters for periods less than 100 days than for greater than 100 days.

\begin{deluxetable}{ccccc}  
\tabletypesize{\small}
\tablecaption{Detection Factor Parameters   \label{table-fdet}}
\tablewidth{0pt}
\tablehead{
  data set &   A     &   B     &    C    &  targets   }
\startdata
    q1-16  &  4.35   &  1.05   &  1.000  &  FGK              \\
    q1-17  &  7.511  &  0.551  &  0.915  &  FGK, p $<$ 100   \\
    q1-17  &  6.93   &  0.83   &  0.83   &  FGK, p $>$ 100
\enddata 
\end{deluxetable}  

An important difference between q1-16 and q1-17 is that the former is believed to have an excess of planets at long periods, and the latter a deficit of these planets (\citet{Coughlin2015}, \citet{Christiansen2016}).  According to these authors this is believed to have occurred because of an incorrect implementation of a ``statistical bootstrap test" which was intended to reduce the number of long-period false positives, but in practice, and after the fact, ``it was found that it also eliminated a significant number of valid long-period, transit-like signals, especially at low signal-to-noise (SNR), which includes many previously designated earth-size, habitable zone planet candidates".  Part of the purpose of the present paper is to compare directly the population estimates from q1-16 and q1-17, to see the practical effect of this data-base difference.  This comparison will help prepare for understanding how to analyze the final data set from Kepler.

\section{The Instrument Model}\label{inst}

By the \textit{instrument model} we mean the expected relationship that connects planets around the Kepler target stars to the observed numbers of detected planets.  In particular, the instrument model algorithm allows us to calculate the expected number of detected planets $N_1(p_i,r_j)$ in Sec.~\ref{intro} for the assumed case of 1 planet per star, uniformly distributed over the full range of $ln(p)$ and $ln(r)$. Then scaling the shape of this array by a least-squares procedure yields the best fit model of the average number of planets per star $\overline{N}$ and the parameters of the distribution function model $h(p,r)$.

Given that we understand in detail many of the properties of the Kepler instrumentation and method of observing (see Sec.~\ref{obs}), this should be relatively straightforward to calculate.  However there are many unknown factors that determine whether a detection can be made, e.g., the exact radius, mass, and temperature of a star, the exact impact parameter, the precise limb-darkening of a given star, the presence of other stars in the image, etc.  In this section we set up the instrument model, using many of the parameters mentioned in Sec.~\ref{obs}, and random numbers for a statistical estimate of the effect of some of the unknowns.  Our method is similar to that used by many authors (e.g., \citet{Mulders2015}) for estimating completeness, but adds more detail regarding limb darkening and the recent work by \citet{Christiansen2016} on detection efficiency.

We start by selecting the range of observed periods ($p_{min}$ to $p_{max}$) and radii ($r_{min}$ to $r_{max}$), and dividing each range into bins ($\Delta \ln(p_i)$ and $\Delta \ln(r_j)$) spaced uniformly in $\ln(p)$ and $\ln(r)$, where $i_{max} = j_{max} = 20$ here.  As discussed in Sec.~\ref{method}, we calculate $N_1$ by assigning 1 planet to each star, which is 1/400 planet per bin, and take the planet's period and radius to be the values at the center of each bin.  For each target star, limited by the ranges of effective temperature and surface gravity as desired, we estimate whether that fractional planet is expected to be detected, and if it is, we add that fraction of a planet, times the probability of detection, to $N_1(p_i,r_j)$.  The probability of detection $P_{det}$ is estimated as follows.


The semi-major axis $a$ is given by
\begin{equation}
  a = \big[ p^2 m_s G/(4\pi^2) \big]^{1/3}
  \label{sma}
\end{equation}
where $m_s$ is the mass of the star, $G$ is the gravitational constant, and we neglect the mass of the planet.  The probability of transit $P_{tr}$ is
\begin{equation}
  P_{tr} = r_s/a
  \label{ptr}
\end{equation}
where $r_s$ is the radius of the star, and we neglect the radius of the planet.
The transit factor $f_{tran}$ is the chord length, relative to the diameter, of a transit at impact parameter $b$,
\begin{equation}
  f_{tran} = (1 - b^2)^{1/2}
  \label{ftr}
\end{equation}
where $b$ is the fraction of a radius from the center to the transit chord.  The impact parameter can in principle be measured from Kepler's transit signature, but the accuracy is not sufficient for present purposes, so we assume that $b$ is given by a random number in the range $(0,1)$.

The transit time $t$ is
\begin{equation}
   t = \frac{r_s \cdot p}{\pi a} \cdot f_{tran} \cdot \sqrt{1 - \overline{e}^2}
  \label{t}
\end{equation}
where the $f_{tran}$ factor converts an equatorial transit time to one at a non-equatorial impact parameter, and the term with the mean eccentricity $\overline{e}$ ($\simeq 0.1$ here) allows for the minor effect of a non-circular orbit \citep{Mulders2015}.

The measurement noise $CDPP$ is found by interpolating to the time of transit $t$ in the table of $CDPP$ values for the target star.  Likewise the MES threshold value is found by interpolating to the time of transit in the table of MES values for the star.

The statistically-expected number of transits $N_{tr}$ in the $span$ of observation of the star is given by
\begin{equation}
  N_{tr} = 1 + int\{ span/p - ran\}
  \label{ntr}
\end{equation}
where $int$ is the integer round-down operator, and $ran$ is a random number in the range $(0,1)$, corresponding to the fact that the start of data-taking occurs at a random point in the first period interval.  Alternatively the value 0.5 can be substituted for $ran$ with about the same statistical result.  In principle the duty-cycle is relevant here, because if that value is small, then there is less chance of the instrument being on duty during a transit, especially for a long-period planet, but since the median duty cycle is close to unity, this term is ignored in the present calculation.  If $N_{tr}$ is less than 3, the planet is deemed not detectable; otherwise the calculation goes to the next step.  The result of this step can be formalized in terms of $f_{num}$ where
\begin{equation}
  f_{num}   =  \left\{
            \begin{array}{ll}
                1       & \qquad  \mbox{for $N_{tr} \geq 3 $ }   \\
                0       & \qquad  \mbox{otherwise}
            \end{array}
            \right.
   \label{eqn-fnum}
\end{equation}

Referring to Appendix~\ref{app-ld}, the limb-darkening coefficient $u$ is interpolated from a table of limb-darkening factors appropriate to the Kepler bandpass, using effective temperature as the interpolating parameter.  The limb-darkening factor $f_{ld}$ is given by
\begin{equation}
  f_{ld} = [1 - u + (u \cdot \pi/4)\cdot (1-b^2)^{1/2}] \Big/ (1 - u/3)
  \label{eqn-fld}
\end{equation}
which gives a value $f_{ld} =1$ for a uniform brightness disk ($u=0$).  As examples, for a typical value of $u=0.5$, $f_{ld}$ ranges from about 1.07 for an equatorial transit ($b=0$) to about 0.67 for a high-latitude transit ($b\simeq 0.99$).

The mission SNR for detecting the planet around the target star is
\begin{equation}
  \mbox{SNR} = (r_p/r_s)^2 \cdot f_{ld} \cdot N_{tr}^{1/2} \Big/ (10^{-6} CDPP(t))
  \label{eqn-snr}
\end{equation}
where $CDPP(t)$ is the tabulated value for the time of transit, and is in units of ppm.

If SNR is less than the MES threshold, then the planet is not detectable, because the star is intrinsically too noisy; otherwise the calculation continues.  The result of this step can be formalized in the $f_{snr}$ term, where
\begin{equation}
  f_{snr}   =  \left\{
            \begin{array}{ll}
                1       & \qquad  \mbox{for $\mbox{SNR} \geq \mbox{MES threshold} $ }   \\
                0       & \qquad  \mbox{otherwise}
            \end{array}
            \right.
   \label{eqn-fsnr}
\end{equation}

The detection efficiency $f_{det}$ of the pipeline has been studied by \citet{Christiansen2015} for q1-16, and by \citet{Christiansen2016} for q1-17, and shown to be representable by an incomplete gamma cumulative distribution function that depends on the mission SNR (or MES) and three empirical parameters $A$, $B$, and $C$, where
\begin{equation}
  f_{det}(z) = \frac{C}{B^A \Gamma(A)} \int_0^z \tau^{A-1} \cdot e^{-\tau/B} \cdot d\tau
  \label{fdet}
\end{equation}
and $z = \mbox{SNR} - 4.1$.  The parameters $A, B, C$ are listed in Table~\ref{table-fdet}.  This functional form is also called the incomplete gamma function (e.g., \citet{Press1992}).

The net result of the instrument calculation is that the probability of detection $P_{det}$ of a planet of period $p_i$ and radius $r_j$ around target star $s_k$ is
\begin{equation}
    P_{det}(p_i, r_j, s_k) = P_{tr} \cdot f_{tran} \cdot f_{num} \cdot f_{snr} \cdot f_{det}
  \label{eqn-probdet}
\end{equation}
as was discussed in Sec.~\ref{method}.

\section{Preparatory Steps}\label{prep}

We can get a rough idea of the distribution function $h(p,r)$ by initially assuming that every star has one planet, equally divided among all bins of $p$ and $r$ under consideration, and comparing the expected number of observed planets from the full mission with the number actually observed.  Any difference will suggest how we should adjust the model.  The steps are as follows.

\begin{figure}[htbp]
\includegraphics[width=3.4in]{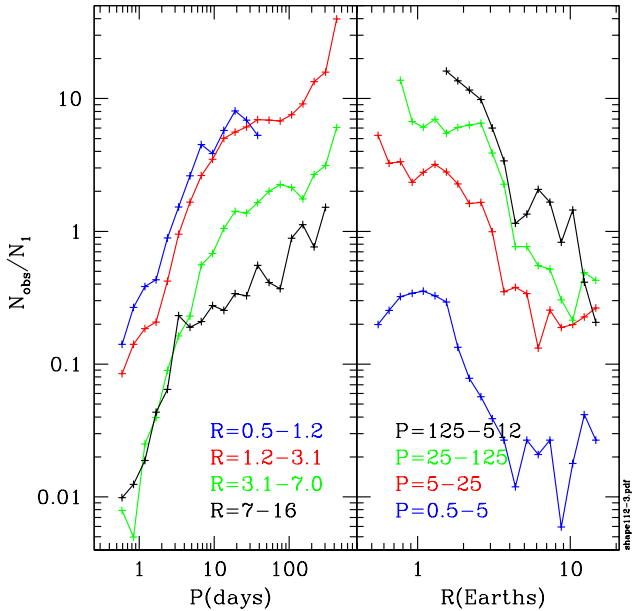}
\caption{ This figure motivates a choice for the functional form of the population of exoplanets with respect to period and radius.  We use the ratio of the observed numbers of planets $N_{obs}$ to the instrumental efficiency $N_1$ in detecting such planets as a proxy for the trend of numbers of planets in the population. On the left, we sum over rows of radius bins from Fig.~\ref{fig-input-planets} to show the broad trend of population numbers as a function of period, where the summing over rows helps in visualizing noisy trends.  For short periods, less than $\sim 8-10$ days, the slopes of curves for all radius groups are similar, and indicate a power-law behavior.  For longer periods the slopes are again similar in magnitude, but shallower, motivating a two-segment power-law model with respect to period. On the right, similar considerations apply to the radius-dependence of the population.  }
\label{fig-shape}
\end{figure}

We put 1 planet in each bin of each target star, and calculate how many of these Kepler will detect in each bin, $N_1(p_i,r_j)$, given the instrument model (above).  This is equivalent to assuming that $h$ is constant.  This procedure generates many planets in almost every bin.  We then form the ratio of the observed number to the simulated number $N_{obs}/N_1$ for each bin.  Some bins have no observed planets, so this ratio is zero for them.  We gauge the trend of this ratio as a function of period by grouping several radius bins for each period bin (to build up the signal to be non-zero), and plotting the result.  Likewise for gauging the trend as a function of radius.  An example is shown in Fig.~\ref{fig-shape} where on the left we show 4 groups of $r$ in order to bring out the $p$-dependence of $h_{approx}$, and on the right we bring out the $r$-dependence.

In the left panel, we see that for all values of radius, the $h_{approx}$ function has a power-law growth with approximately a single slope for $p < 6 \pm 2$, and a shallower but still roughly constant slope for larger $p$ values. In the right panel we see a similar power law behavior, with a roughly constant value of slope for $r < 2.0 \pm 0.5$, and a steeper slope for large values of $r$.  In addition, the separation of the curves suggests that the underlying $h(p,r)$ function might be representable by a product of $p$ and $r$ power laws.

The simplest separable power law would be to use a single power for each of the full ranges of $p$ and $r$.  It is clear from the breaks in Fig.~\ref{fig-shape} that this will not be successful, but it is instructive to try it nevertheless.  Thus we write
\begin{equation}
  h_{single}(p,r) = f(p) \cdot g(r)
  \label{eqn-hfg}
\end{equation}
where each function is separately normalized, giving
\begin{eqnarray}
  \int_{p_{min}}^{p_{max}} f(p) \cdot d\ln(p) & = & 1   \label{eqn-fnorm}\\
  \int_{r_{min}}^{r_{max}} g(r) \cdot d\ln(r) & = & 1   \label{eqn-gnorm}
\end{eqnarray}
which is in agreement with Eqn.~\ref{eqn-h}.
Explicitly the $f$ and $g$ functions can be written
\begin{eqnarray}
  f(p) & = & \beta \cdot p^b    \label{eqn-fsingle} \\
  g(r) & = & \alpha \cdot r^a   \label{eqn-gsingle}
\end{eqnarray}
where the powers $a$ and $b$ are the key fitting parameters, and the coefficients $\alpha$ and $\beta$ are determined by Eqns.~\ref{eqn-fnorm} and \ref{eqn-gnorm}.

\begin{figure}[htbp]
\centering
\includegraphics[width=3.4in]{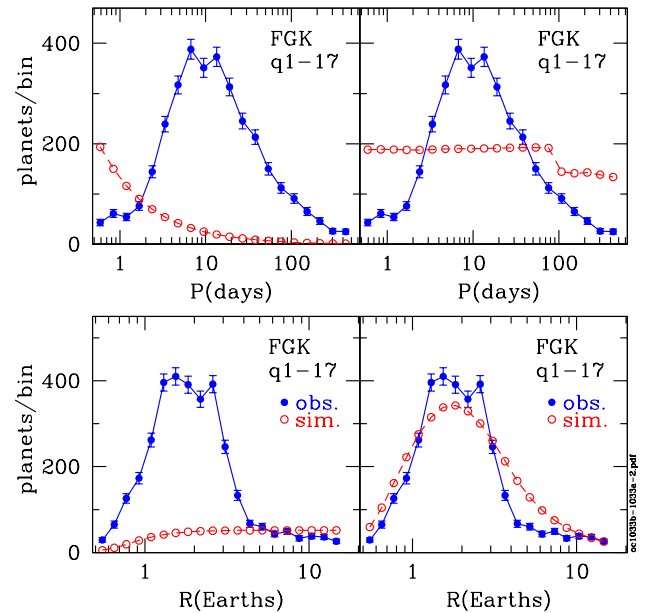}
\caption{ (left) The q1-17 data (blue points) are summed to show the period and radius variations explicitly, and the best-fit models for the case of a simulated flat population in $\ln(r)$ and $\ln(p)$ space, i.e., the powers of $p$ and $r$ are $a = 0$ and $b = 0$, respectively. The fall-off at long periods and small radii reflects the low instrument efficiency at these extremes. (right) For the same data (q1-17) here the powers are allowed to vary, but are constant over the full ranges of $p$ and $r$.  Already, for this very simple model, the radius variation is beginning to match the observed data.  And the period variation is at least coming closer to matching the data.  These panels provide motivation for the two-segment model, which will allow the small and large radius slopes to be modeled independently, as well as the short and long period slopes. }
\label{fig-fgk-zero-const}
\end{figure}

We now consider all $20\times20$ bins of $N_{obs}(p,r)$, and carry out the least-squares fit in Eqns.~\ref{eqn-chisq1} and \ref{eqn-chisq2}.  We consider the simplest possible case, with flat distributions in log space given by $a = 0$ and $b = 0$.  The parameter vector then has a single element: $\overrightarrow{v} = \{\overline{N} \}$.  The result of this fit is shown in Fig.~\ref{fig-fgk-zero-const} (left), where it is clear that the fit is very poor, but for good reasons.  For the period plot, the simulated number of detected planets falls rapidly for long periods, because the instrument efficiency falls off roughly as $p^{-1}$, so long-period planets are simply not detected as well as short-period ones.  For the radius plot, the instrument efficiency is low for small planets, owing to their low SNR, so small planets are less detectable than large ones, and if all sizes are equally likely in the model population, as is the case here, then the trend in the lower plot will be realized.

For the next case we allow $a$ and $b$ to become non-zero, so the parameter vector is  $\overrightarrow{v} = \{a, b, \overline{N}\}$.  The result of this fit is shown in Fig.~\ref{fig-fgk-zero-const} (right), where we find the $p$ distribution is better, but still a poor approximation.  However the $r$ distribution is surprisingly good, considering the simplicity of the simulation, giving a peaked distribution with very roughly the correct width.  These encouraging fits have $a = -1.57$ and $b = 0.98$, powers that are in rough agreement with the directions and magnitudes of the slopes in Fig.~\ref{fig-shape}.  From a physical point of view, the $b \approx 1$ power of $p$ puts increasing numbers of planets per log interval at larger periods, approximately counteracting the above-mentioned fall-off in instrument sensitivity, resulting in a nearly flat distribution of expected detections.  Note also that now a small step appears at $p \sim 100$ days, a result of the estimated efficiency drop for long periods noted by \citet{Christiansen2015}.  For the radius plot, the negative power of $r$ ($a \approx -1.6$) in this population model forces the number of large planets to be small, thus allowing small planets to be the dominant ones that are detected, although still dominated by the instrumental fall-off for very small planets, with the net result that we get a peaked distribution of detected planets, limited on the small-$r$ end by the instrument and on the large-$r$ end by the population itself.

Returning to our observation that Fig.~\ref{fig-shape} strongly suggests that a broken power law is needed, we pursue this type of population model in the following section.

\section{Broken Power-Law Model} \label{broken}

The demonstration in Fig.~\ref{fig-shape} that the inferred population of planets appears to be dominated by a broken power law in each of $p$ and $r$ suggests that we try distribution functions of the form

\begin{equation}
  f(p)  =  \left\{
            \begin{array}{ll}
                \beta_1 \cdot p^{b_1}       & \qquad  \mbox{for $p < p_1$ }   \\
                \beta_2 \cdot p^{b_2}       & \qquad  \mbox{otherwise}
            \end{array}
            \right.
   \label{eqn-broken-1}
\end{equation}
and
\begin{equation}
  g(r)  =  \left\{
            \begin{array}{ll}
                \alpha_1 \cdot r^{a_1}       & \qquad  \mbox{for $r < r_1$ }   \\
                \alpha_2 \cdot r^{a_2}       & \qquad  \mbox{otherwise.}
            \end{array}
            \right.
   \label{eqn-broken-2}
\end{equation}
The parameter vector now has 7 terms, $\overrightarrow{v} = \{\overline{N}, p_1, r_1, b_1, b_2, a_1,a_2 \}$, and the coefficients are determined by requiring continuity of the segments
\begin{eqnarray}
   \beta_1  \cdot p_1^{b_1} & = &  \beta_2  \cdot p_1^{b_2}    \label{eqn-cont1}  \\
   \alpha_1 \cdot r_1^{a_1} & = &  \alpha_2 \cdot r_1^{a_2}    \label{eqn-cont2}
\end{eqnarray}
as well as normalization to unity (Eqns.~\ref{eqn-fnorm} and \ref{eqn-gnorm}).  An explicit solution to these 4 equations is given in Appendix~\ref{app-norm} to illustrate the procedure.

We apply this broken power law to data from q1-16 as well as q1-17 in the following.  We show results for FGK stars considered together, because this gives the highest SNR results owing to the large number of planets.  We also apply the model to each of F, G, K, and M stars, to search for trends with stellar temperature or mass, and to have a look forward in anticipation of the final data release from the Kepler project.  The results are summarized in Tables~\ref{table-q1-16} and \ref{table-q1-17}.

In each case we limit the analysis to the 2-segment power laws for each of $p$ and $r$, although it will become clear that 3-segment laws could probably do a better job of fitting the data.  Since the final data are not yet available, it is prudent at this stage to keep the analysis relatively simple, and search instead for trends and lessons.

The fitting procedure is to assume a trial value of the $r_1$ and $p_1$ break points, guided by Fig.~\ref{fig-shape}, then run the least-squares fitting program to solve for the remaining 5 parameters $\overrightarrow{v} = \{\overline{N}, b_1, b_2, a_1,a_2 \}$, and iterate this procedure on the break points to achieve an overall minimum chi-square value.  The resulting $(p_1, r_1)$ values are added as a footnote to the tables.

We find that the least-squares fit needs to use uniform weights rather than Poisson weights. One reason is that there are many bins in which there are just 0  observed planets, so it is not clear how to assign a $\sqrt{n}$ value for the $n=0$ case.  Another reason is that trial runs with Poisson weights, including a weight of 1 for the $n=0$ bins, gave results that systematically failed to generate enough planets to match the bins with up to 50 or so planets, and systematically failed to match the total number of observed planets, probably owing to the large number of bins with 0 or 1 planets.  However we find that using uniform weights always succeeds in giving a number of simulated planets that is close to the total observed number, so this method is used throughout.

The error bars reported for each quantity in Tables~\ref{table-q1-16} and \ref{table-q1-17} are obtained by choosing the larger of two values: (1) the uncertainty reported by the least-squares fitting program; or (2) the maximum deviation of a parameter obtained by randomly dividing the set of observed planets into 2 groups, each with about half the total number of planets, then independently fitting each group, and selecting the maximum deviation of each parameter as the quoted error.  (For the final Kepler data set, another method could be to choose the RMS of parameters resulting from many runs with random half-data sets, but in the present paper we use the simpler method above.)

\section{Results 1: Occurrence of Spectral Types}  \label{occ-spty}

In this section we apply the broken power-law model to the q1-16 and q1-17 data sets, and look at the results as a function of spectral type.  In the second section we look at the same results but from the perspective of the derived parameters $\overline{N}, a_1, a_2, b_1, b_2$.  The third section examines how the results relate to the question of the occurrence of Earth-like planets.

\begin{figure}[htbp]
\centering
\includegraphics[width=3.4in]{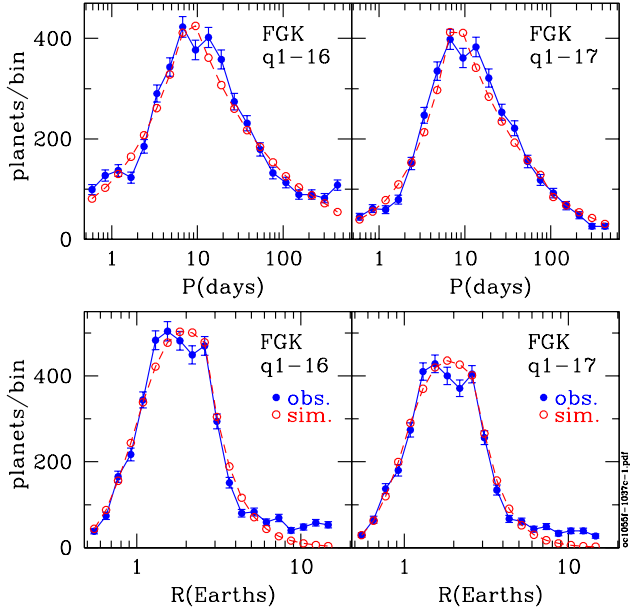}
\caption{ (left) The observed numbers of planets around FGK stars for q1-16 are moderately well-matched by a simple two-segment power law in each parameter.  The most significant lack of a good fit is for radii greater than about 5 Earth radii, where the need for an additional power law is indicated, but is not pursued in this paper.  (right) A similar comparison of data and simulation for the q1-17 data set.  Note the differences with respect to the left panel at the shortest and longest periods. }
\label{fig-fgk-q1-16-17}
\end{figure}

\subsection{FGK Stars} \label{fgk-stars}

The results of the 2-segment power-law fitting procedure, as applied to the group of FGK stars, are shown in Fig.~\ref{fig-fgk-q1-16-17} for q1-16 (left) and q1-17 (right).  Recall that the model is fit to the full $20\times 20$ grid of bins of observed planets, but we are displaying the net result of summing along rows or columns to produce the summary plots in these figures.  We draw several conclusions from comparing these two figures.

(1) Overall the $p$ and $r$ fits look encouragingly good, although not perfect.  This is an initial validation of both the segmented power-law approach, as well as the separability of the $p$ and $r$ functional forms. Further validation should of course be carried out when the final data set is available.

(2) The largest relative discrepancies occur for large planet radii, $r > 6$ Earth radii, where the observed number of planets continues to be flat and finite, but the model continues its downward slope toward zero.  This area is a candidate for a 3rd segment in the model for the next data set if this feature persists.

(3) The next-largest lack of a good fit is in the peaks of the distributions, where by eye it appears that the observed data have a flat plateau over about a factor of 2 in both period and radius.  The fit may improve after a 3rd segment is added to the $r$ data, and possibly the $p$ data as well, but at present it looks like a modest-width plateau is indicated.  And if this is the case, it remains to be seen if this is a result of a change in slope of the parent population, or a subtlety of the instrument model.

(4) Interestingly, there are large differences in the shape of the $p$ data sets at small as well as large periods.  The q1-16 data has 2-3 times more planets than the q1-17 data in the 0.5 to 1.5 day range and also in the 200 to 500 day range, although at the peak of the distributions both data sets are nearly the same.  There is no such obvious shape change for the $r$ data between q1-16 and q1-17.  Thus the issue of false positives for q1-16 and false negatives for q1-17, mentioned at the end of Sec.~\ref{obs}, appears to be confined to the extremes of period, and does not seem to favor any particular range of radius.

A summary of the fitting results for FGK and the individual spectral types is shown in the top line of Tables~\ref{table-q1-16} and \ref{table-q1-17}.  The key result from these fits is that the average number of planets per FGK star is about $\overline{N}(FGK) \simeq 4.9 \pm 0.3$ planets per star, in the range of radii from 0.5 to 16 Earth radii and period from 0.5 to 512 days.  This value may change when the final data set and corresponding analysis becomes available.

\begin{figure}[htbp]
\centering
\includegraphics[width=3.4in]{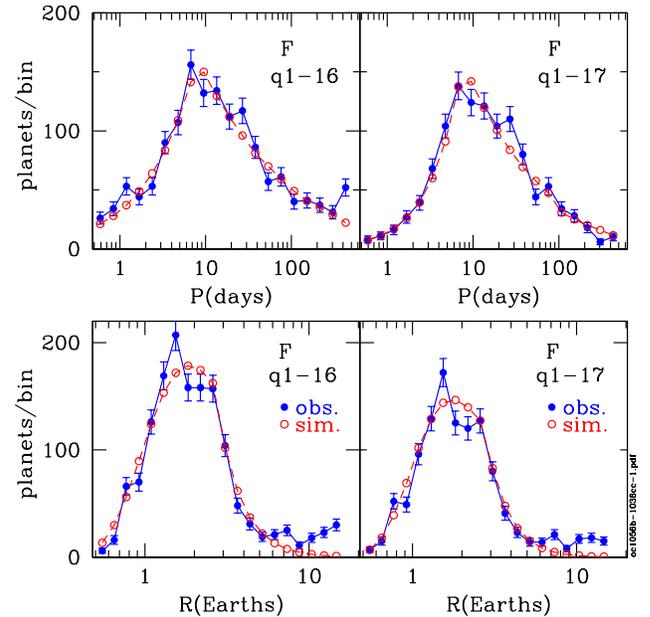}
\caption{ (left) Similar to Fig.~\ref{fig-fgk-q1-16-17}, but here for F stars only.  Similar conclusions apply. }
\label{fig-f-q1-16-17}
\end{figure}

\begin{deluxetable*}{ccccccccc}  
\tabletypesize{\small}
\tablecaption{Fit results q1-16   \label{table-q1-16}}
\tablewidth{6in}
\tablehead{
   SpTy & $p_1$ & $r_1$ & $\overline{N}$    &  $a_1$             &  $a_2$              &  $b_1$            &  $b_2$  }
\startdata
   FGK  & 8.0  & 2.6   & $ 4.70 \pm 0.23 $ & $ -0.68 \pm 0.04 $ & $  -2.86 \pm 0.04 $ & $ 1.52 \pm 0.03 $ & $ 0.43 \pm 0.02 $  \\
    F   & 8.0  & 2.6   & $ 5.36 \pm 0.73 $ & $ -0.94 \pm 0.14 $ & $  -3.03 \pm 0.14 $ & $ 1.67 \pm 0.05 $ & $ 0.52 \pm 0.02 $  \\
    G   & 8.0  & 2.6   & $ 4.82 \pm 0.12 $ & $ -0.67 \pm 0.04 $ & $  -2.43 \pm 0.09 $ & $ 1.51 \pm 0.05 $ & $ 0.41 \pm 0.01 $  \\
    K   & 8.0  & 2.6   & $ 3.93 \pm 0.18 $ & $ -0.21 \pm 0.11 $ & $  -3.72 \pm 0.21 $ & $ 1.37 \pm 0.05 $ & $ 0.28 \pm 0.03 $  \\
    M   & 11.5 & 2.3   & $ 5.14 \pm 1.56 $ & $  0.42 \pm 0.55 $ & $  -7.98 \pm 4.50 $ & $ 1.41 \pm 0.11 $ & $-0.18 \pm 0.18 $
\enddata 
\end{deluxetable*}  

\subsection{F Stars} \label{f-stars}

Repeating the analysis for the F-star groups gives the results shown in Fig.~\ref{fig-f-q1-16-17} for q1-16 (left) and q1-17 (right).  The overall shapes of the $p$ and $r$ distributions of observed planets appear to be similar to the FGK shapes.  The resulting average number of planets per F star is $\overline{N}(F) \simeq 5.7 \pm 0.8$, larger than the FGK group value, but only at about the $\sim 1 \sigma$ level.

\begin{deluxetable*}{ccccccccc}  
\tabletypesize{\small}
\tablecaption{Fit results q1-17   \label{table-q1-17}}
\tablewidth{6in}
\tablehead{
   SpTy & $p_1$ & $r_1$ & $\overline{N}$ &  $a_1$        &  $a_2$          &  $b_1$        &  $b_2$  }
\startdata
   FGK  & 7.6 & 2.7  & $ 5.04 \pm 0.23 $ & $ -0.81 \pm 0.02 $ & $ -3.22 \pm 0.17 $  & $ 1.85 \pm 0.05 $ & $  0.39 \pm 0.03 $  \\
    F   & 7.6 & 2.7  & $ 5.95 \pm 0.88 $ & $ -1.15 \pm 0.12 $ & $ -3.39 \pm 0.16 $  & $ 2.15 \pm 0.04 $ & $  0.48 \pm 0.03 $  \\
    G   & 7.6 & 2.7  & $ 5.13 \pm 0.26 $ & $ -0.77 \pm 0.05 $ & $ -2.74 \pm 0.23 $  & $ 1.79 \pm 0.06 $ & $  0.38 \pm 0.05 $  \\
    K   & 7.6 & 2.7  & $ 4.09 \pm 0.32 $ & $ -0.31 \pm 0.19 $ & $ -4.33 \pm 0.29 $  & $ 1.62 \pm 0.04 $ & $  0.24 \pm 0.03 $  \\
    M   & 11.5 & 2.2 & $ 7.77 \pm 1.77 $ & $ -0.14 \pm 0.35 $ & $-12.55 \pm 7.88 $  & $ 1.55 \pm 0.15 $ & $  0.13 \pm 0.30 $
\enddata 
\end{deluxetable*}  

\begin{figure}[htbp]
\centering
\includegraphics[width=3.4in]{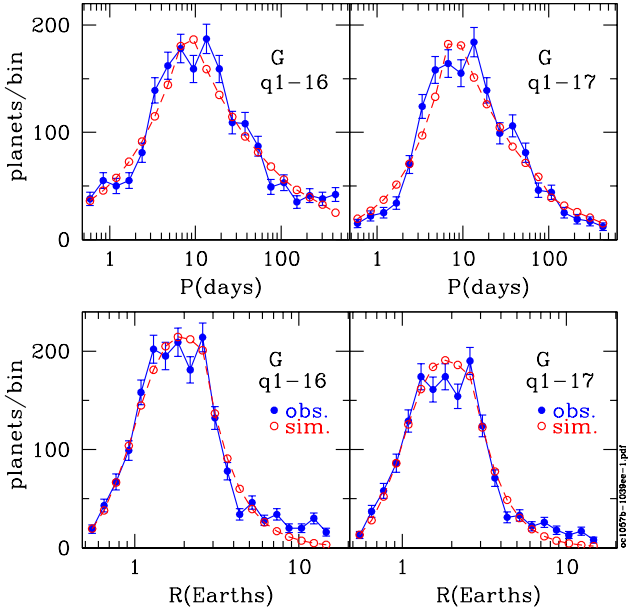}
\caption{ Similar to Fig.~\ref{fig-fgk-q1-16-17}, but here for G stars only.  Similar conclusions apply. }
\label{fig-g-q1-16-17}
\end{figure}

\subsection{G Stars} \label{g-stars}

The fitting results for G stars are shown in Fig.~\ref{fig-g-q1-16-17} for q1-16 (left) and q1-17 (right).  The shapes of the $p$ and $r$ distributions are very similar to those for F stars. The average number of planets per G star is $\overline{N}(G) \simeq 5.0 \pm 0.2$, fewer than for F stars, but with a smaller uncertainty.

\begin{figure}[htbp]
\centering
\includegraphics[width=3.4in]{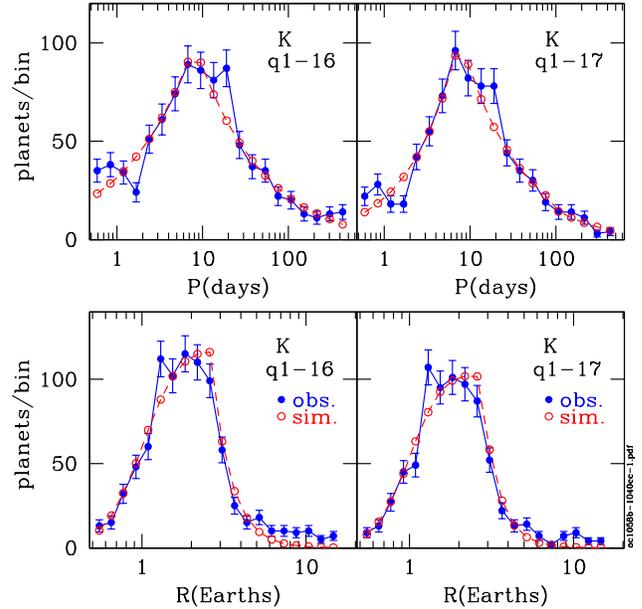}
\caption{ Similar to Fig.~\ref{fig-fgk-q1-16-17}, but here for K stars only.  Similar conclusions apply. }
\label{fig-k-q1-16-17}
\end{figure}

\subsection{K Stars} \label{k-stars}

The results for K stars are shown in Fig.~\ref{fig-k-q1-16-17} for q1-16 (left) and q1-17 (right).  The shapes of the $p$ and $r$ distributions are again similar to those for F and G stars. The average number of planets per K star is $\overline{N}(K) \simeq 4.0 \pm 0.3$, fewer than for F or G stars.  The $\overline{N}$ values now form a trend, at slightly better than the $1 \sigma$ level, with more planets being around hotter and more massive stars.  This trend is worth looking at more carefully when the final data set becomes available.

\begin{figure}[htbp]
\centering
\includegraphics[width=3.4in]{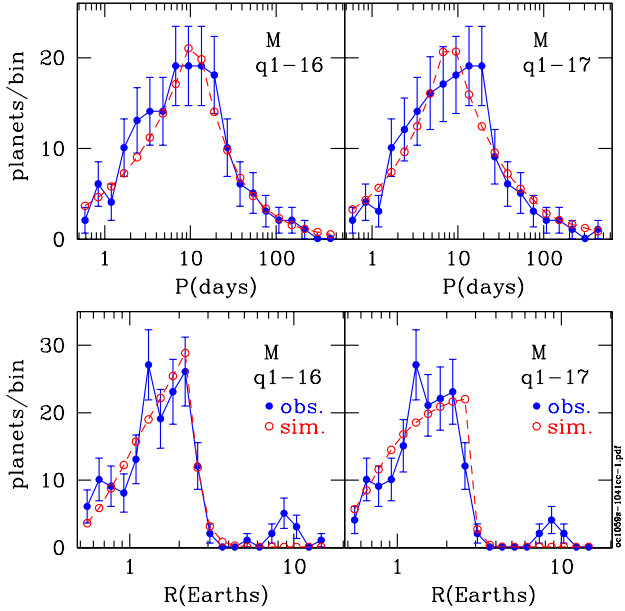}
\caption{ Similar to Fig.~\ref{fig-fgk-q1-16-17}, but for M stars only.  Here the comparison is made difficult by the small numbers of M stars in the Kepler sample, giving poor statistical results.  Another difficulty is the apparent change in character of the distribution of observed planets, cutting off very steeply for radii greater than about 2 Earth radii, and showing an isolated small peak around 9 Earth radii.  It is also possible that the efficiency model for M stars is not the same as for FGK stars, which if true reduces the comparitive value of the derived parameters in these data sets. }
\label{fig-m-q1-16-17}
\end{figure}

\subsection{M Stars} \label{m-stars}

The fitting results for M stars are shown in Fig.~\ref{fig-m-q1-16-17} for q1-16 (left) and q1-17 (right).  The shape of the $p$ distribution is again roughly similar to the F, G, and K distributions.  But the $r$ shape shows clear signs of being different, with relatively more observed planets at small radii, and fewer planets in the region around $r \simeq 3$ Earth radii.  The sharp drop in the simulated number of observed planets at this point can be seen in the large negative power of $r$, the $a_2$ term in Tables~\ref{table-q1-16} and \ref{table-q1-17}.

 The average number of planets per M star is $\overline{N}(M) \simeq 6.5 \pm 1.7$, which is roughly 2 times larger than might be expected if the trend from F to K is simply extrapolated.  It is not clear whether this large number of planets per star is a real feature of the population, or if it is an artifact of the very small numbers of M stars in the sample.  It is worth noting that M-star transit observations are especially susceptible to contamination from background stars, owing to their relative faintness, a potential problem for all types of transits \citep{Ciardi2015}.

\begin{figure}[htbp]
\centering
\includegraphics[width=3.4in]{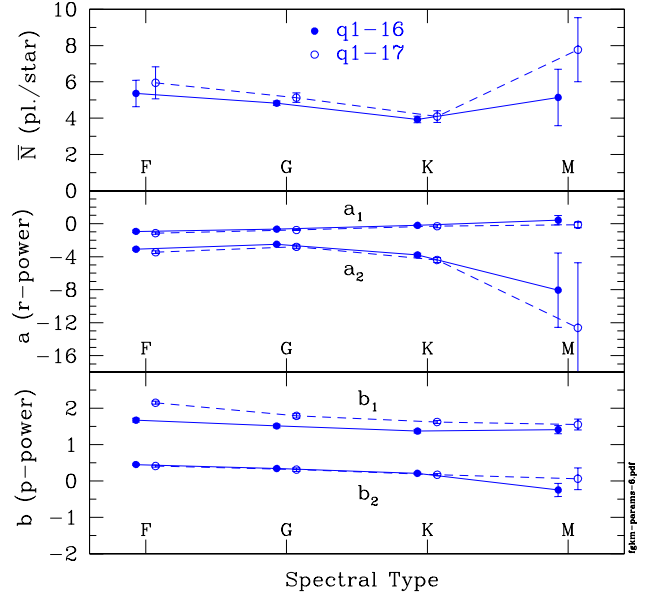}
\caption{ This figure summarizes the derived parameters of $h(p,r)$ for the planet population from q1-16 and q1-17 for each spectral type of host star.  (top) The average number of planets per star  in the simulated population, $\overline{N}$, is shown as a function of spectral type, F to M.  For F, G, and K stars this value is about 5 planets per star.  But for M stars it jumps to a larger value, which may be real, but may also be a function of the small-numbers statistics, as well as an incorrect model of instrumental efficiency for observing M stars.   (middle) The power $a$ of radius in the model is shown, where $a_1$ applies to small radii, and $a_2$ to large radii.  (bottom)  The power $b$ of period in the model is shown, where $b_1$ applies to short periods, and $b_2$ to long periods. }
\label{fig-params}
\end{figure}

\section{Results 2: Occurrence Parameters $\overline{N}, a_1, a_2, b_1, b_2$}  \label{occ-param}

The physical meaning of each of the fitted $\overline{N}$, $a$, and $b$ parameters is worth examining.  We plot these parameters in Fig.~\ref{fig-params}. In most cases there is a clear trend with spectral type, which is likely a clue to the origin and evolution of these systems, and is certainly of interest when estimating the science yield of future exoplanet missions.

\subsection{Average number of planets per star} \label{disc-nbar}

The average number of planets per star, $\overline{N}$, is shown in Fig.~\ref{fig-params} (top) for each of F-M spectral types.  As mentioned in Sec.~\ref{broken}, the F-G-K trend in $\overline{N}$ runs from about 5.7 to 4.0 planets per star, suggesting that an extrapolation might give a value of about 3.1 for M stars.  However the observed value is about 6.5, roughly twice the extrapolated value, about a $2 \sigma$ deviation.  So a key question is which of these values, if any, is closest to the actual population value.  We shall see in the following subsections that the trend of each $a$ and $b$ parameter is nominally in line with extrapolation for the F-G-K parameters, so the $\overline{N}(M)$ one stands alone in this regard.

The value of $\overline{N}$ for G stars is about $5.0 \pm 0.2$ in the present analysis, for the range of $p$ and $r$ considered.  The $p$ range in the Solar System includes Mercury, Venus, and Earth, and reaches about half-way to Mars, so the corresponding Solar System value might be taken to be about 3.5.  For any individual system the $1 \sigma$ range is $\overline{N} \pm \sqrt{\overline{N}}$ which runs from 2.8 to 7.2, so on this basis the number of planets in this part of the Solar System is clearly within the expected range.  Also the relatively steep fall-off of population with radius (the $r^{a_2}$ dependence) suggests that giant planets are rare, on average, so it is not a surprise that there are none in the Solar System out to nearly Mars.

\subsection{Planet Frequency vs Radius}  \label{disc-a1,2}

The frequency of planets varies as $r^a$, where the $a$ values for smaller ($a_1$) and larger ($a_2$) planets are listed in Tables~\ref{table-q1-16} and \ref{table-q1-17} and shown in Fig.~\ref{fig-params} (middle).  Recall that the population $N(p,r)$ varies as $\partial N/\partial \ln(r) \sim r^a$ or $\partial N/\partial r \sim r^{a-1}$.  The $a_1$ parameter varies smoothly from about -1 for F stars to about 0 for M stars, telling us that the population of small planets $(r < 2.6)$ varies as $\partial N/\partial r \sim r^{-2}$ for F stars, but $r^{-1}$ for M stars.  In both cases the population has more planets per star with small radii than large (up to $r \sim 2.6$).  In other words, the steepening effect is more dramatic for F stars than M stars.

For large planets with $r \sim 2.6$ to $5$ Earth radii, the $a_2$ parameter tells us that  $\partial N/\partial r \sim r^{-4}$ for F and G stars, and  $\partial N/\partial r \sim r^{-9}$ or steeper for M stars, with K stars in between these extremes.  So in all cases the drop-off toward larger radii is very steep.  However this effect occurs only within the range about 2.6 to 5 Earth radii, and it is much flatter for larger radii.  See, for example, Fig.~\ref{fig-fgk-q1-16-17} where the fitted curve departs from the observed numbers of planets at about the $r \sim 5$ point, beyond which this paper does not explicitly model the data.

\subsection{Planet Period Variation with Spectral Type}  \label{disc-b1,2}

The frequency of planets varies as $p^b$, where the $b$ values for short- ($b_1$) and long-period ($b_2$) planets are listed in Tables~\ref{table-q1-16} and \ref{table-q1-17} and shown in Fig.~\ref{fig-params} (bottom).  For FGK stars the occurrence increases steeply with period as $\partial N/\partial \ln(p) \sim p^{1.7}$, averaging the $b_1$ values for FGK, from $p = 0.5$ to $\sim 8$ days period.  Here the $b_1$ values are systematically smaller for the q1-16 data than q1-17, because the q1-16 data have a relatively large number of planets at very short periods in q1-16 compared to q1-17.

At long periods, greater than about 8-12 days, the frequency of planets increases more slowly with period as $\partial N/\partial \ln(p) \sim p^{0.4}$ for FGK stars.  However at the longest periods, about 150 to 500 days, the differences between q1-16 and q1-17 are again dramatically different, as here the q1-16 numbers of observed planets increases (Fig.~\ref{fig-fgk-q1-16-17}) whereas for q1-17 it decreases.  The $b_2$ parameter fit does not fully reflect this difference because the fit is based on the larger number of points between about 8 and 150 days.  However $b_1$ or its equivalent is crucial for estimating the frequency of habitable-zone planets, which is hampered here by the lack of agreement between q1-16 and q1-17 in this period range.

The long-period slope of occurrence is systematically flatter for late-type stars, varying as
$\partial N/\partial \ln(p) \sim p^{0.5}$, for F stars and trending smoothly to $\sim p^{0.3}$ for K stars, indicating that the F-star systems tend to have more long-period planets than K stars.

\section{Results 3: Occurrence of Earth-Like Planets}  \label{occ-eta123}

\subsection{Eta Earth} \label{eta-earth}

To facilitate the characterization and comparison of various exoplanet population estimates, three index terms are in common use.  The first term is $\eta_{\oplus}$, usually (but not always) defined as
\begin{equation}
  \eta_{\oplus} = \sharp \; \mbox{terr. planets per star, in HZ}
  \label{eqn-eta-def}
\end{equation}
where here \textit{terrestrial} will be (somewhat arbitrarily) taken to be the radius range $r_l = 0.5$ to $r_u = 1.25$ Earth radii, and \textit{habitable zone} (HZ) will be (similarly) taken to be the the nominal range 0.80 to 1.80 AU around the Sun, as recommended by \citet{Kasting2012} in \citet{Traub2012}.  This translates to an insolation range of 1.563 to 0.309 times that of the present Earth, a range that can be applied to orbits around any star with a known effective temperature and radius.  The usefulness of $\eta_{\oplus}$ is that it relates to the zone where water could conceivably be liquid on a planet's surface for any type of star, but the downside is that it requires a consensus on the definitions of terrestrial and habitable zone, both of which are widely debated in the literature.  The values in this paper are for illustration, and are not intended to be definitive definitions.

We incorporate these definitions by projecting out the relevant part of the $h(p,r)$ distribution function using projection factors $f_{terr}$ and $f_{HZ}$ in the following integral,
\begin{eqnarray}
  \eta_{\oplus}(SpTy) & = & \overline{N} \cdot \int_{terr} \int_{HZ} f_{terr}(r) \cdot f_{HZ}(p,SpTy)  \nonumber \\
                      &   &  \cdot h(p,r) \cdot d\ln(p) \cdot  d\ln(r),
  \label{eqn-eta}
\end{eqnarray}
which is adapted directly from Eqn.~\ref{eqn-n}.
The radius factor is then
\begin{equation}
  f_{terr}(r)  =  \left\{
            \begin{array}{ll}
                1       & \qquad  \mbox{for $r_l < r < r_u$ }   \\
                0       & \qquad  \mbox{otherwise.}
            \end{array}
            \right.
   \label{eqn-fterr}
\end{equation}

The period factor is estimated by examining every star in the Kepler target star list, selecting the desired subset of stars of a given spectral type, and allowing each star to be surrounded by a series of planets in circular orbits with periods ranging from $p_{min}$ to $p_{max}$.  For each such planet, we ask if the insolation from the star at the planet's orbital distance is in the HZ range.  We count the number $\Delta N(p,SpTy,HZ)$ of such planets in a given period range $\Delta p$, and we count the total number of target stars that were considered $\Delta N(p,SpTy)$.  Then the period factor is given by the fraction of stars that could have a planet in the HZ at each possible period,
\begin{equation}
  f_{HZ}(p_i,SpTy) =  \frac{\Delta N(p_i,SpTy,HZ)}{\Delta N(p_i,SpTy)},
  \label{eqn-fhz}
\end{equation}
where here the period $p$ is taken to be the mean period in each bin under consideration.

\begin{figure}[htbp]
\centering
\includegraphics[width=3.4in]{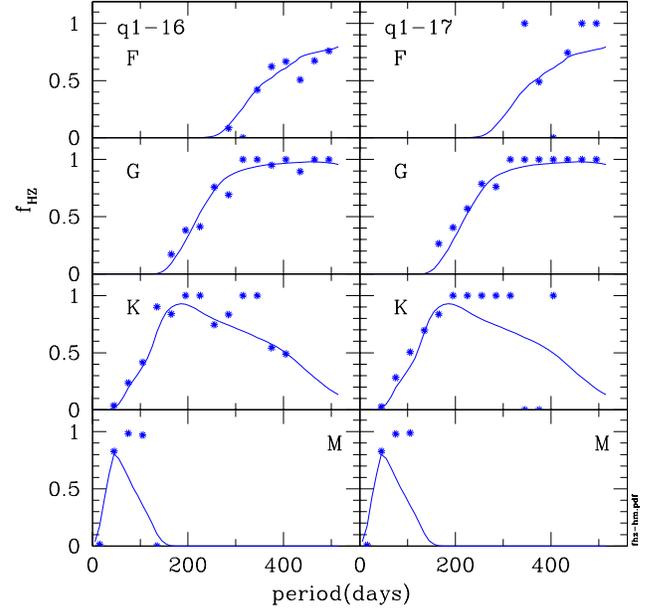}
\caption{ The period projection factor $f_{HZ}(p)$ is shown as a function of period for each data set, q1-16 and q1-17, and for each spectral type.  The smooth curves show the fraction of target stars that can have a HZ planet at the indicated period.  The points show the fraction of observed planets that fall in the HZ, which owing to the much smaller number of samples gives a more highly scattered indication of $f_{HZ}$.  See text for details.}
\label{fig-fhz}
\end{figure}

The fraction $f_{HZ}(p_i,SpTy)$ is shown as a smooth line in Fig.~\ref{fig-fhz} for each spectral type and each set of data.  The target star lists for q1-16 and q1-17 are so similar that the curves for each data set differ by much less than $1\%$ at nearly all periods.  As expected, for hot stars the HZ occurs at long periods, and for cool stars it peaks at shorter periods.  The curves for F and G stars are truncated by the artificial limit of considering only periods less than $p_{max}$; obviously they could be extended to longer periods, but for the purpose of the present paper they are limited as shown.  For K and M stars the $f_{HZ}$ factor is clearly well-contained within the Kepler period range, so estimates of $\eta_{\oplus}$ for these stars will not be limited by the Kepler mission length.

The radius range in Eqn.~\ref{eqn-fterr} lies wholly within the model radius segment $r < r_1$.  The period range in Eqn.~\ref{eqn-fhz} likewise lies within the period segment $p > p_1$.  Therefore Eqn.~\ref{eqn-eta} can be explicitly written as
\begin{eqnarray}
  \eta_{\oplus}(SpTy) &  = &  \overline{N} \cdot \Bigg[ \alpha_1 \cdot \Big(\frac{r_u^{a_1} - r_l^{a_1}}{a_1}\Big) \Bigg] \nonumber \\
  &  &  \cdot \Bigg[ \sum_{p_i} \beta_2 \cdot \Big( \frac{p_{i,u}^{b_2} - p_{i,l}^{b_2}}{b_2} \Big) \cdot f_{HZ}(p_i) \Bigg],
  \label{eqn-eta2}
\end{eqnarray}
where the range limits $(r_l, r_u)$ and $(p_l, p_u)$ are given above, and the sum over $p_i$ is in terms of the bin sizes chosen for the smooth curve in Fig.~\ref{fig-fhz}.  The estimated values of $\eta_{\oplus}(SpTy)$ are listed in Table~\ref{table-eta}.

As a cross check, a similar procedure was applied to the observed planets, and shown as single points.  The numbers of observed planets are much smaller than the numbers of target stars, so these data are much noisier, and in some cases the $\Delta N(p_i,SpTy,HZ)$  number is equal to $\Delta N(p_i,SpTy)$ so the ratio is unity.  Most error bars would extend well off the plot range so they are not shown.  These points are included here merely to roughly corroborate the $f_{HZ}$ curve from stars.

The F-star values of $\eta$ are almost certainly underestimated, because the $f_{HZ}$ curve is truncated at 512 days, and clearly there is a significant range beyond that point where $f_{HZ}$ remains large.  Likewise the extrapolated value of $f_{HZ}(p)$ beyond 512 days may well remain large, although this determination awaits the final Kepler data release.  Thus $\eta_{\oplus}(F)$ will certainly be larger than the tabulated values of 0.59 and 0.66.

The G-star values of $\eta$, close to unity in the Table, are likewise almost certainly underestimated, but by less than the F-star values.  The $f_{HZ}$ factor will turn over beyond 512 days, and the extrapolation of $f_{HZ}(p)$ will depend on the final Kepler data, but clearly $\eta_{\oplus}(G)$ will be larger than the average of about 1.0 in the Table.

The K-star values of $\eta$ are probably the best-determined of the group here, because the number of samples is large enough that the error bars are relatively small, and the $f_{HZ}$ factor is well-contained within the Kepler period range.  On this basis, the average value of $\eta_{\oplus}(K) \simeq 0.73$ is likely to hold up well in the next data release.  A second conclusion from the G and K values is that the $\eta$ value appears to be dropping toward cooler stars; confirmation of this trend must await the final Kepler data release.

The M-star value of $\eta$ is very uncertain owing to the small number of M stars and planets in the Kepler database.  This uncertainty also shows up in the disparity in value between the two data sets.  Thus from the data at hand, it is not possible to say whether or not the $\eta$ value for M stars is large, although on the basis of the two values shown, the $\eta$ value does appear to be larger than might be expected from a simple extrapolation of the G and K values.

\subsection{Gamma Earth} \label{gamma-earth}

The second term is $\Gamma_{\oplus}$, here taken to be the average number of planets per star in unit intervals of $\ln(p)$ and $\ln(r)$, where $p$ and $r$ are in units of days and Earth radii, evaluated at the present Earth, i.e.,
 \begin{eqnarray}
  \Gamma_{\oplus} & = & \frac{d^2N}{d\ln(p) \cdot d\ln(r)} \Bigg|_{p_{\oplus},r_{\oplus}} \\
                  & = & \overline{N} \cdot h(p_{\oplus}, r_{\oplus})                      \\
                  & = & \overline{N} \alpha_1 \beta_2 \cdot r_{\oplus}^{a_1} \cdot p_{\oplus}^{b_2}
  \label{eqn-gamma}
\end{eqnarray}
For perspective, the effective range in $p$ and $r$ implied by the $\Delta\ln(p) =\Delta\ln(r) = 1$ interval, if split evenly in log space, corresponds to upper and lower values of $p_u/\overline{p} = e^{1/2} \simeq 1.65$ and $p_l/\overline{p} = e^{-1/2} \simeq 0.61$, and likewise for $r$.  The corresponding radius range is roughly 0.6 to 1.6 Earth radii, and period range is 220 to 600 days (compare Venus at 225 days, and Mars at 687 days).

The usefulness of $\Gamma_{\oplus}$ is that it allows direct comparison of various estimates of occurrence without the complication of specifying a numerical range of period or radius.  A disadvantage is that for some spectral types, M for example, there are essentially no planets near the $(p_{\oplus}, r_{\oplus})$ point.  (A solution to this disadvantage would be to re-define $\Gamma_{\oplus}$ as centered at $(I_{\oplus}, r_{\oplus})$ where $I_{\oplus}$ is planet insolation in units of the current insolation at Earth.)

\begin{deluxetable}{ccccc}  
\tabletypesize{\small}
\tablecaption{Measures of occurrence: $\eta_{\oplus}$ and $\Gamma_{\oplus}$   \label{table-eta}}
\tablewidth{0pt}
\tablehead{
   SpTy & $\eta_{\oplus}$(q1-16) &  $\eta_{\oplus}$(q1-17) & $\Gamma_{\oplus}$(q1-16) & $\Gamma_{\oplus}$(q1-17)  }
\startdata
    F   & $ 0.59 \pm 0.12 $    &    $ 0.66 \pm 0.14 $      &      -                  &       -                \\
    G   & $ 0.97 \pm 0.02 $    &    $ 1.03 \pm 0.10 $      &   $ 1.10 \pm 0.19 $     &    $ 1.14 \pm 0.94 $   \\
    K   & $ 0.72 \pm 0.02 $    &    $ 0.75 \pm 0.11 $      &      -                  &       -                \\
    M   & $ 0.75 \pm 0.33 $    &    $ 1.23 \pm 0.18 $      &      -                  &       -
\enddata 
\end{deluxetable}  

Evaluating $\Gamma$ for the q1-16 and q1-17 data sets, using the parameters in Tables~\ref{table-q1-16} and \ref{table-q1-17}, we find $\Gamma_{\oplus} = 1.10 \pm 0.19$ and $1.14 \pm 0.94$ respectively, for an average of $1.1 \pm 0.2$, listed in Table~\ref{table-eta}.

By comparison, \citet{Foreman2014} find $\Gamma_{\oplus} = 0.019^{+0.019}_{-0.010} $ which is clearly not in agreement, by a large factor.  The same reference quotes their interpretation of \citet{Petigura2013b} as finding  $\Gamma_{\oplus} = 0.119^{+0.046}_{-0.035} $, which is likewise inconsistent.  We have no ready explanation for these widely divergent values.

\subsection{Zeta Earth} \label{zeta-earth}

The third term is $\zeta_{1.0}$, introduced by \citet{Burke2015}, and defined as the number of planets per star within $\epsilon = \pm 20\%$ of $p_{\oplus}$ and $r_{\oplus}$.  This is a variant of $\Gamma_{\oplus}$, as can be seen by using Eqn.~\ref{eqn-n} and evaluating the result with $\epsilon \ll 1$, giving
\begin{equation}
  \zeta_{1.0} \simeq 4 \epsilon^2 \Gamma_{\oplus} \simeq 0.16 \cdot \Gamma_{\oplus}.
  \label{eqn-zeta}
\end{equation}
We find $\Gamma_{\oplus} = 1.1 \pm 0.1$ which gives $\zeta_{\oplus} = 0.18 \pm 0.02$, in rough agreement with \citet{Burke2015} who find $\zeta_{\oplus} = 0.1$, with a range of 0.01 to 2.0.

\section{Summary}  \label{summary}

The two most recent Kepler data releases on the numbers of transiting exoplanets are clearly converging to a stable set of values in the broad sense, however each of these releases has known internal flaws that make them unreliable for drawing conclusions on the detailed dependence of the exoplanet population on the basic parameters from Kepler: planet period and radius, and stellar spectral type.  Nevertheless, we have shown here that broad characteristics of the population can be inferred from the present data, and it is expected that the method of analysis introduced here will be of value for analyzing the final data set.


\acknowledgments

Acknowledgements.  I thank the Kepler Team for providing such spectacularly abundant and precise data.
I thank the staff at the Computation Facility of the Harvard-Smithsonian Center for Astrophysics for advice and computer facilities.  I thank Rachel Akeson, Geoff Bryden, Jessie Christiansen, John Krist, and Bill Press for advice and assistance.  I acknowledge helpful conversations with Tom Barclay, Chris Burke, Jeff Coughlin, Jon Jenkins, and Gijs Mulders.  Part of this research was carried out at the Jet Propulsion Laboratory, California Institute of Technology, under a contract with the National Aeronautics and Space Administration.
This research has made use of the NASA Exoplanet Archive, which is operated by the California Institute of Technology, under contract with the National Aeronautics and Space Administration under the Exoplanet Exploration Program.

Copyright 2016 California Institute of Technology.  Government sponsorship acknowledged.







\appendix
\vspace{0.2in}

\section{A.  Limb Darkening}\label{app-ld}

The apparent disk of a star at visible wavelengths is usually darker near the limb than at the center (e.g., \citet{Claret2011}); exceptions include the near-infrared region and rapid-rotators.  This visible limb-darkening means that as a Kepler planet traverses a chord across a limb-darkened star, it first enters at a relatively dark limb, then crosses a brighter part symmetrically disposed about the meridian, and finally exits across a dark limb.  If $\theta$ is the angle between radius vectors from the center of the spherical star to the observer (Kepler) and a point on the surface of the star, and if $\rho$ is the distance from the center of the apparent disk to a point on the disk, in units of the stellar radius $R_s$, then, letting $\mu = \cos\theta$, and assuming a linear limb-darkening rule, the flux per unit area $I(\rho)$ at any point on the apparent disk is given by
\begin{equation}
  I(\rho) = I(0) \cdot [1 - u(1-\mu(\rho))]
  \label{ld1}
\end{equation}
where $I(0)$ is the flux per unit area at the apparent center of the disk, and $u$ is the linear limb-darkening parameter.  Integrating this over the disk gives the full-disk flux $F_s$ as
\begin{equation}
  F_s = 2\pi \int_0^1 I(\rho) \cdot (R_s \rho) \cdot d(R_s \rho) .
  \label{ld2}
\end{equation}
Substituting, and evaluating the integral, we get the full-disk flux as
\begin{equation}
  F_s = \pi R_s^2 \cdot I(0) \cdot (1 - u/3)
  \label{ld3}
\end{equation}
which is the product of the area of the star, times the flux per unit area at the center, times a limb-darkening correction which in general gives a smaller full-disk flux than would be expected in the absence of limb-darkening.  For the present simulations, specific values of $u$, for the case of the Kepler bandpass, are taken from Fig.~1 in \citet{Claret2011}.  The limb-darkening parameter and effective temperature pairs $(u, T_{eff}(K))$ are as follows: (0.60, 3200), (0.60, 3700), (0.74, 4400), (0.67, 5000), (0.56, 6000), (0.49, 7000), and (0.40, 10000).  Linear interpolation is used to obtain $u$ from $T_{eff}$.

During a transit, the planet can cross the disk at any chord across that disk.  We represent this by using an impact parameter $b$, where $b$ is a random number in the range $(0,1)$.  In an $(x,y)$ coordinate system on the disk, the transit path is parallel to the $x$ axis, and crosses the disk at a distance $y = b \cdot R_s$ perpendicular to the $x$ axis.  The corresponding time for a single transit is $t_1$, given by
\begin{equation}
  t_1 = (2 R_s / v) \cdot f_{tran}
  \label{ld4}
\end{equation}
where $v$ is the velocity of the planet in a circular orbit, and
\begin{equation}
  f_{tran} = \sqrt{1 - b^2}
  \label{ld5}
\end{equation}
is the transit factor.  The $x$-coordinate of the chord extends from $+x_{tran}$ to $-x_{tran}$, where
\begin{equation}
  x_{tran} = R_s f_{tran} .
  \label{ld6}
\end{equation}
During a transit the planet blocks an instantaneous flux $F_p$ from the star intensity $I(\rho)$ (flux per unit area) at that position, giving
\begin{equation}
  F_p(x) = I(\rho(x)) \cdot \pi R_p^2  .
  \label{ld7}
\end{equation}
The average blocked flux $\overline{F_p}$ during the transit is
\begin{equation}
  \overline{F_p} = \int_{-x_{tran}}^{+x_{tran}} F_p(x) dx  \Bigg/  \int_{-x_{tran}}^{+x_{tran}} dx  .
  \label{ld8}
\end{equation}
Evaluating this expression gives
\begin{equation}
  \overline{F_p} = F_s \cdot (R_p/R_s)^2 \cdot f_{LD}
  \label{ld9}
\end{equation}
which is simply the averaged star flux, times the fractional area of the star which is blocked by the planet, times a limb-darkening factor $f_{LD}$ which is given by
\begin{equation}
  f_{LD} = [ 1 - u + (u\pi/4) \cdot f_{tran} ] \Big/ (1 - u/3),
  \label{ld10}
\end{equation}
where $f_{LD}$ is on the order of unity, but can be greater than or less than 1.0 depending upon the value of the impact parameter and the limb darkening parameter.  Note that $f_{LD} = 1.0$ for an impact parameter $b_0 = 0.529$, independent of the value of $u$.

The signal $S_1$ for a single transit is given by
\begin{equation}
  S_1 = \overline{F_p} t_1 = F_s t_1 (R_p/R_s)^2 f_{LD}
  \label{ld11}
\end{equation}
which is the same equation as for an equatorial transit but modified by a shorter transit time for all values of $b$, and a limb-darkening factor which will be greater than unity for a transit closer to the equator than $b_0$, or less than unity for a transit farther from the equator than $b_0$.  Furthermore, averaging Eqn.~\ref{ld11} over all values of $b$, from 0 to 1, gives the average value as
\begin{eqnarray}
  \overline{S_1} & = & \int_0^1 S_1(b) db  \Bigg/  \int_0^1 db                 \\
                 & = & S_1(b=0) \cdot \pi/4 .            \label{ld12}
\end{eqnarray}
So the average signal from a planet transit is always less than the signal from an equatorial transit, for all values of $u$, including when $u=0$ in an equatorial transit across an uniform-brightness star.  (Note that $\pi/4$ is also simply the ratio of the area of a circular disk to the area of a square star, where using only the equatorial chord essentially amounts to assuming a square star.)  Thus, taking into account a range of impact parameters $b$, we will get an average transit signal that is $\pi/4$ or about 0.79 times the signal from an equatorial transit, for any value of the limb-darkening parameter $u$.

Another way to say this is that the ensemble-average planet transit signal in a simulation with random impact-parameter transits of limb-darkened stars will give the same result as the ensemble average planet signal in a simulation with equatorial transits of uniform-brightness stars, where the uniform-brightness stars have their radii reduced by a factor 0.79.

Yet another way to view this is to say that a simulation with equatorial transits of uniform-brightness stars will give a population of planets that is at least 0.79 times too small, compared to a simulation that includes random impact parameters, even before taking into account the increased noise of a shorter transit.

\vspace{0.2in}
\section{B.  Power-Law Normalization}\label{app-norm}

For the \textit{single-segment} power laws given by Eqns.~\ref{eqn-fsingle} and \ref{eqn-gsingle}, subject to the normalization requirement in Eqns.~\ref{eqn-fnorm} and \ref{eqn-gnorm}, and assuming that the least-squares fit is designed to adjust the powers $b$ and $a$, the coefficients are given by
\begin{eqnarray}
  \beta  & = & \Bigg[ \frac{p_{max}^b - p_{min}^b}{b} \Bigg]^{-1}  \label{eqn-beta}  \\
  \alpha & = & \Bigg[ \frac{r_{max}^a - r_{min}^a}{a} \Bigg]^{-1}  \label{eqn-alpha}
\end{eqnarray}
which fully determines the distribution functions $f(p)$ and $g(r)$, and therefore $h(p,r) = f \cdot g$.

For the \textit{two-segment} power laws given by Eqns.~\ref{eqn-broken-1} and \ref{eqn-broken-2}, subject to the normalization in Eqns.~\ref{eqn-fnorm} and \ref{eqn-gnorm}, and the continuity requirement in Eqns.~\ref{eqn-cont1} and \ref{eqn-cont2}, the $\beta$ coefficients are given by
\begin{eqnarray}
  \beta_1  & = &  \Bigg[ \frac{p_1^{b_1} - p_{min}^{b_1}}{b_1} + \Bigg(\frac{p_1^{b_1}}{p_1^{b_2}}\Bigg) \frac{p_{max}^{b_2} -p_1^{b_2}}{b_2} \Bigg]^{-1}  \label{eqn-beta1} \\
  \beta_2  & = &  \beta_1 \cdot \frac{p_1^{b_1}}{p_1^{b_2}}                   \label{eqn-beta2}  \\
\end{eqnarray}
and similarly the $\alpha$ coefficients are
\begin{eqnarray}
  \alpha_1  & = &  \Bigg[ \frac{r_1^{a_1} - r_{min}^{a_1}}{a_1} + \Bigg(\frac{r_1^{a_1}}{r_1^{a_2}}\Bigg) \frac{r_{max}^{a_2} -r_1^{a_2}}{a_2} \Bigg]^{-1}  \label{eqn-alpha1} \\
  \alpha_2  & = &  \alpha_1 \cdot \frac{r_1^{a_1}}{r_1^{a_2}} .                  \label{eqn-alpha2}  \\
\end{eqnarray}
If $b$ or $a$ are close to zero, then for numerical accuracy it may be prudent to replace $p_1^b/b$ or $r_1^a/a$ with $\ln(p_1)$ or $\ln(r_1)$.




\clearpage   

\clearpage





\end{document}